\documentclass[prd,amsmath,amssymb,superscriptaddress,twocolumn,nofootinbib,floatfix]{revtex4}
\input{epsf}
\usepackage{epsf}
\usepackage{graphicx,epsfig}
\usepackage{bm}
\usepackage{latexsym,float}
\usepackage{color}

\newcommand {\ga} {\ {\raise-.5ex\hbox{$\buildrel>\over\sim$}}\ }
\newcommand {\la} {\ {\raise-.5ex\hbox{$\buildrel<\over\sim$}}\ }
\newcommand{\eqn}[1] {Eq.~(\ref{#1})}
\newcommand{\fig}[1] {Fig.~(\ref{#1})}
\newcommand{\phid}{\dot{\phi}}

\def\be{\begin{equation}}
\def\ee{\end{equation}}
\def\ba{\begin{eqnarray}}
\def\ea{\end{eqnarray}}
\renewcommand{\(}{\left(}
\renewcommand{\)}{\right)}

\renewcommand{\]}{\right]}

\begin{document}

\title{Unparticle dark energy}

\author{De-Chang Dai} \email{ddai@buffalo.edu}
\affiliation{HEPCOS, Department of Physics, SUNY at Buffalo, Buffalo, NY 14260-1500}

\author{Sourish~Dutta}
\email{sourish.d@gmail.com}
\affiliation{Department of Physics and Astronomy, Vanderbilt University,
Nashville, TN 37235}

\author{Dejan Stojkovic} \email{ds77@buffalo.edu}
\affiliation{HEPCOS, Department of Physics, SUNY at Buffalo, Buffalo, NY 14260-1500}

\begin{abstract}
We examine a dark energy model where a scalar unparticle degree of freedom plays the role of quintessence. In particular, we study a model where the unparticle degree of freedom has a standard kinetic term and a simple mass potential, the evolution is slowly rolling and the field value is of the order of the unparticle energy scale ($\lambda_u$). We study how the evolution of $w$ depends on the parameters $B$ (a function of the unparticle scaling dimension $d_u$), the initial value of the field $\phi_i$ (or equivalently, $\lambda_u$) and the present matter density $\Omega_{m0}$.  We use observational data from Type Ia supernovae, baryon acoustic oscillations and the cosmic microwave background to constrain the model parameters and find that these models are not ruled out by the observational data. From a theoretical point of view, an unparticle dark energy model is very attractive, since unparticles (being bound states of fundamental fermions) are protected from radiative corrections. Further, coupling of unparticles to the standard model fields can be arbitrarily suppressed by raising the fundamental energy scale $M_F$, making the unparticle dark energy model free of most of the problems that plague conventional scalar field quintessence models.
\end{abstract}

\maketitle

\section{Introduction}

Cosmological data from a wide range of sources including type Ia supernovae \cite{union08, perivol, hicken}, the cosmic microwave background \cite{Komatsu}, baryon acoustic oscillations \cite{Seo:2005ys,Percival:2007yw}, cluster gas fractions \cite{Samushia2007,Ettori} and gamma ray bursts \cite{Wang,Samushia2009} seem to indicate that at
least $70\%$ of the energy density in the
universe is in the form of an exotic, negative-pressure component,
called dark energy.  (See Ref. \cite{Copeland} for a recent
review). It is usually assumed that a scalar field with a postulated potential may yield the equation of state needed to drive the acceleration. While it is easy to come up with such a potential (see e.g. \cite{RatraPeebles,Caldwell:1997ii,SteinhardtWangZlatev,Stojkovic:2007dw,dutta}), the underlying problem is that we have not discovered a single fundamental scalar field so far. Fundamental scalar fields, being unprotected from radiative corrections, are very problematic since they require a significant fine tuning in theory (see e.g.  \cite{Carroll,HsuMurray}). Couplings of the quintessence field with other standard model particles, in particular fermions, must be suppressed  \cite{Brax:1999yv,Doran:2002bc,Arbey:2007vu,BasteroGil:2009eb} (however, see \cite{Frieman,Dutta,Albrecht} for a discussion of PNGB quintessence models which are radiatively stable  and see \cite{DuttaHsu,Stojkovic:2007dw}
for a quintessence model which can couple strongly enough with standard model particles to be detected in colliders). It is therefore instructive to search for other degrees of freedom which may effectively play a role similar to the scalar field.

Recently, the existence of a new scale invariant sector very weakly coupled to the standard model was postulated \cite{Georgi:2007ek,Georgi:2007si,Georgi:2008pq}. The fundamental energy scale, $M_F$, of this sector is perhaps far beyond the reach of today's or near-future accelerators. However, the existence of such a sector may affect low energy phenomenology. The effective low energy field theory which describes these effects is often called unparticle physics since these new degrees of freedom would not behave as ordinary particles. For example, their scaling dimension does not have to be an integer or half an integer. Because of the scale invariance, the fundamental particles are massless.  An unparticle is a composite state of the fundamental massless particles, and couples to the standard model particles through a heavy mediator. This heavy mediator has a mass of the order of  $M_F$. Thus, interactions between the unparticles and the standard model particles is suppressed by powers of $M_F$.
Another important characteristic of unparticles is that their mass  can take continuous values \cite{Nikolic:2008ax}. For example, the mass of an unparticle with energy $\omega$ can take all the values from zero to $\omega$. A lot of work \cite{Zhang:2008zzy,He:2008xv,He:2008ef,Kumar:2008zzb,Kikuchi:2008pr,Kikuchi:2008nm} recently focused on the new collider signals for unparticle physics.
The unparticles could also play the role in gravity and cosmology \cite{Boyanovsky:2008bf,Davoudiasl:2007jr,Wei:2008nc,Lee:2009ny,
Chen:2009ui,Chen:2009ys,Kikuchi:2007az,Liao:2007ic,Goldberg:2008zz,Mureika:2007nc,Bertolami:2009qn,Bertolami:2009jq}. Black hole Hawking radiation of unparticles was studied in  \cite{Dai:2008qn}.

In this paper we will try to answer the question whether scalar unparticle degrees of freedom can play the role of dark energy. An unparticle dark energy model is interesting for a number of reasons. In the simple model of unparticles constructed in \cite{Georgi:2007si}, the fundamental degrees of freedom at high energies are fermions. Unparticles are bound states of these fermions, but at energies below a certain  ``phase transition scale" $\lambda_u$ they appear as the fundamental degrees of freedom. This situation is analogous to quarks and mesons in QCD. While fundamental scalar fields are not protected from radiative corrections, unparticles are, since (like in technicolor models) the fundamental fields are fermions whose mass is protected.
Also, coupling of unparticles to the standard model fields can be arbitrarily suppressed by raising the fundamental energy scale $M_F$. Therefore, the unparticle dark energy model is free of most of the problems that plague fundamental scalar field quintessence models.

In exploring this idea further, we will study in detail an unparticle dark energy model with $\phi^2$ quintessence potential. Since the effective scalar unparticle degree of freedom $\phi$ is quite different from the fundamental scalar (i.e. different scaling dimension, continuous mass etc.) the phenomenology may be quite different from the standard one. We will derive the effective functional dependence of the energy density of the universe on $\phi$ and $\dot{\phi}$ in order to calculate the Hubble expansion rate as the function of the same parameters. We will then confront the type Ia supernova (SNIa), Baryon Acoustic Oscillation (BAO) and Cosmic Microwave Background (CMB) data with the predictions of the model.

\section{Unparticle scalar field}

Consider a scalar unparticle degree of freedom $\phi$, with the standard kinetic term and simple potential $V(\phi) \sim \phi^2$.
The energy density and pressure can be written as
\begin{eqnarray}
\label{rp}
\rho_u &=& \frac{1}{2}\left(\frac{\partial \phi}{\partial t}\right)^2 + \frac{M^2\phi^2}{2}\\
P_u &=& \frac{1}{2}\left(\frac{\partial \phi}{\partial t}\right)^2 - \frac{M^2\phi^2}{2}
\end{eqnarray}
where $M$ is the mass of this field. Since unparticles do not have a fixed mass, we take the mass $M$ to be the average of the mass distribution
\begin{equation} \label{M}
M^2= \frac{\int_0^{\lambda_u} \rho (\mu)\mu^2d\mu^2}{\int_0^{\lambda_u} \rho (\mu)d\mu^2}
\end{equation}
where $\rho (\mu)$ is the mass spectral density of unparticles, while $\lambda_u$ is the scale of the phase transition above which unparticle description stops being valid and therefore is the maximum mass of unparticles. It is important to distinguish $\lambda_u$ from $M_F$. While $M_F$ is the new fundamental scale with ordinary particle degree of freedom, $\lambda_u$ is the scale at which these fundamental degrees of freedom are seen as unparticles.
The normalization factor is introduced to preserve the unparticle number. We believe that taking the average of the (continuous) unparticle mass distribution is the most convenient way to capture an average collective influence of unparticles on the evolution of the energy density of the universe. Of course, this is not a unique way to do this, one may for example consider two fields, one at the top and one at the bottom of the mass distribution. It will be interesting to consider alternative approaches, however, we take here the minimalistic average behavior approach.
%It is interesting that $\lambda_u$ depends on the energy density of %unparticles, i.e. it is large when the energy density is large and %small when the energy density is small, which opens the possibility %to address both primordial inflation and late time acceleration of %the universe.

The mass spectral density of unparticles with continuous mass \cite{Dai:2008qn} can be written as
\begin{equation}\label{massd}
\rho(\mu)=\frac{\mu^{2d_u-4}}{\Lambda^{2d_u-2}_u}
\end{equation}
where $d_u$ is the scaling dimension of unparticles. The scale  $\Lambda_u$ is included for dimensional reasons and in the simplest case is equal to the ``phase transition scale" $\lambda_u$. Strictly speaking the exact normalization of unparticles is unknown since we do not know how many unparticle degrees of freedom there are, neither we know the precise combinatorics factors (see discussion in \cite{Dai:2008qn}). However, since we are taking an average of the mass distribution of unparticles in Eq.~(\ref{M}), precise normalization of unparticles does not matter. Substituting (\ref{massd}) in (\ref{M}) we get
\begin{equation} \label{M1}
M^2=\frac{d_u-1}{d_u}\lambda_u^2
\end{equation}

We consider now an unparticle scalar field  with negligible kinetic energy, which is an analog of a slowly rolling scalar field. From Eq. (\ref{rp}), the energy density is
\begin{equation}
\label{density-lambda}
\rho_u =\frac{d_u-1}{2d_u}\lambda_u^2\phi^2
\end{equation}

Now, we make the assumption that, as long as the kinetic energy is negligible, $\lambda_u \approx \phi$. This is just a statement that the value of the field is of the same order of magnitude as the maximum unparticle mass for the relevant energy
scale. In this case we can write equation (\ref{density-lambda}) as
\begin{equation}
\label{rho-lambda}
\rho_u \sim \frac{d_u-1}{2d_u}\lambda_u^4
\end{equation}
We can now express the mass term as
\begin{equation}
\frac{1}{2}M^2 \equiv \frac{d_u-1}{2d_u}\lambda_u^2=  B \sqrt{\rho_u} \, .
\end{equation}

The constant $B$ is, in general, a function of the unparticle scaling dimension $d_u$. If we set the constant of proportionality between $\lambda_u$ and $\phi$ to unity, we obtain
\be
\label{Bdu}
B= \sqrt{\frac{d_u-1}{2d_u}}
\ee
While this may not be an exact expression for $B$, it should be correct up to an $O(1)$ factor.

The energy density and pressure can now be replaced by
\begin{eqnarray} \label{rhoold}
\rho_u &=& \frac{1}{2}\left(\frac{\partial \phi}{\partial t}\right)^2 + B \sqrt{\rho_u}\phi^2\\
P_u &=& \frac{1}{2}\left(\frac{\partial \phi}{\partial t}\right)^2 - B \sqrt{\rho_u}\phi^2
\end{eqnarray}

In what follows, we will use this unparticle setup in the cosmological context. We will assume that the scalar unparticle degree of freedom plays the role of a quintessence field.

We note here that in \cite{Chen:2007qc} a different equation of state was derived for unparticle degrees of freedom. However, there, a model of bosons in thermal equilibrium was used, which is appropriate for the dark matter description, but not for our purpose of a slowly rolling scalar field.

\section{The Model}

We first rewrite the relation \eqn{rhoold} for our purpose as  (dots denote time derivatives):
\be
\label{rho_useless}
\rho_u=\frac{1}{2}\dot{\phi}^2+B\sqrt{\rho_u}\phi^2+\rho_\Lambda
\ee
where we have shifted the potential by the cosmological constant $\rho_\Lambda$ for convenience.
From here, one can express $\rho_u$ as
\be
\label{rho}
\rho_u=\frac{1}{4}\(B\phi^2+\sqrt{B^2\phi^4+2\dot{\phi}^2+4\rho_\Lambda}\)^2
\ee

The potential function of the scalar field $V\(\phi,\dot{\phi}\)$ is therefore
\be
\label{V}
V\(\phi,\phid\)=\frac{1}{2}\(B^2\phi^2+B\sqrt{B^2\phi^4+2\phid^2+4\rho_\Lambda}\)\phi^2+\rho_\Lambda
\ee

The Lagrangian of the unparticle field is:
\be
\label{Lagrangian}
{\cal L}=\frac12\(\partial_\mu \phi\)^2-V\(\phi,\phid\)
\ee

From the Euler-Lagrange equations, we obtain the following equation of motion for $\phi$ (where a subscripted variable denotes a partial derivative with respect to that variable):
\be
\label{phi eom}
\ddot{\phi}\(1-V_{\phid\phid}\)+\(3H-V_{\phi\phid}\)\phid+V_{\phi}-3HV_{\phid}=0
\ee

Now consider a Universe consisting of perfect fluid dark matter, radiation and unparticle dark energy. The dark matter energy density $\rho_m$ and the radiation density $\rho_r$ satisfy the continuity equations
\be
\label{rhom eom}
\dot{\rho}_m+3H\rho_m=0
\ee

\be
\label{rhor eom}
\dot{\rho}_r+4H\rho_r=0
\ee
The evolution of the expansion rate $H\equiv\frac{\dot{a}}{a}$ (where $a$ is the scale factor) is given by the Friedmann equation
\be
\label{Friedman}
H^2=\frac{8\pi G}{3}\(\rho_m+\frac{1}{2}\phid^2+V\(\phi,\phid\)\)
\ee

The system of equations \eqref{phi eom}-\eqref{Friedman} is solved numerically. Initial conditions are set at some arbitrary redshift $z_i$ deep within the radiation dominated era. For the numerical analysis we work in units where $c=\hbar=8\pi G=1$, and we scale our time coordinate such that the present day Hubble value $H_0=1$ for a $\Lambda CDM$ cosmology (in what follows, $0$-subscripts will always denote present day values). The initial value of the matter and radiation densities can be obtained by appropriately redshifting the observed present-day matter and radiation densities to $z_i$. The initial velocity of the field is set to zero, since the field can be assumed to be frozen on account of the very high Hubble friction - we have checked that even if the field is given an initial velocity it quickly comes to rest. The evolution is continued until the matter density parameter reaches the observed value of $\Omega_{m 0}\approx 0.3$.

Clearly, the three parameters that completely determine the evolution of the system are the following:
\begin{itemize}
\item The observed present matter density parameter $\Omega_{m,0}$.
\item The constant $B$, which is a function of the scaling dimension of unparticles, as explained above (see \eqn{Bdu} and the discussion below it).
\item The initial value of the field $\phi_i$. As per our assumptions that $\phi\sim\lambda_u$, if the kinetic term is negligible, and that the field is slowly rolling, the parameter $\phi_i$ is essentially the same as initial $\lambda_u$.
\end{itemize}

\begin{figure}
	\epsfig{file=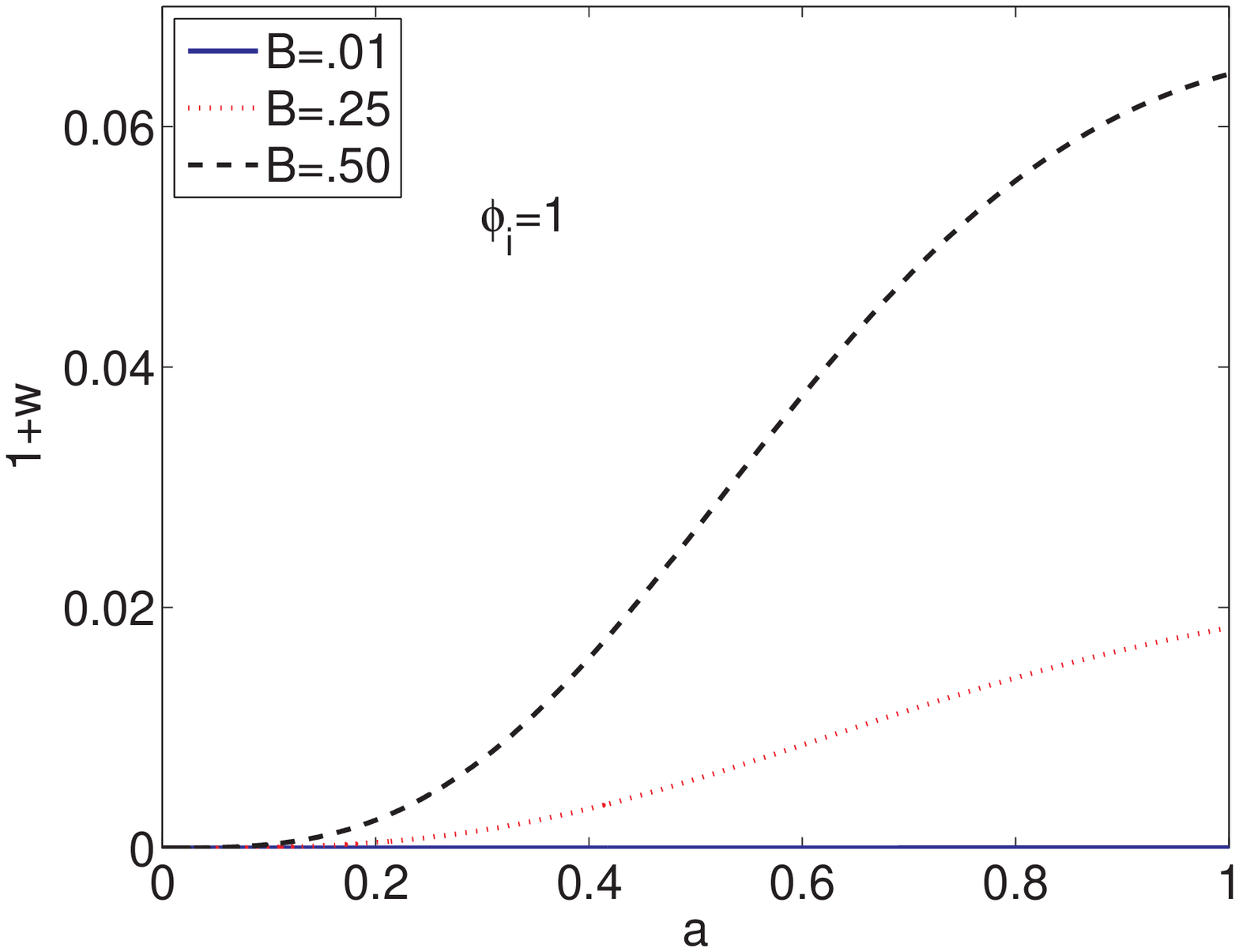,height=51mm}
	\caption
	{	\label{onepw_vs_a_phi1} $1+w(a)$ for different choices of $B$. $\phi_i$ is fixed to $1$}
\end{figure}

We now consider the evolution of the equation of state of the unparticle dark energy. The effect of varying $B$ on the function $w(a)$ is shown in \fig{onepw_vs_a_phi1}, where the initial value of the field is fixed at $\phi_i=1$. Note that $w$ always starts out at $-1$, as a result of our choosing the initial velocity of the field to be zero. For small values of $B$, $w$ increases slowly. However, as $B$ increases, $w$ evolves faster. If one chooses a smaller fixed value of $\phi_i$, the plots have similar shapes, but smaller amplitudes.

\begin{figure}
	\epsfig{file=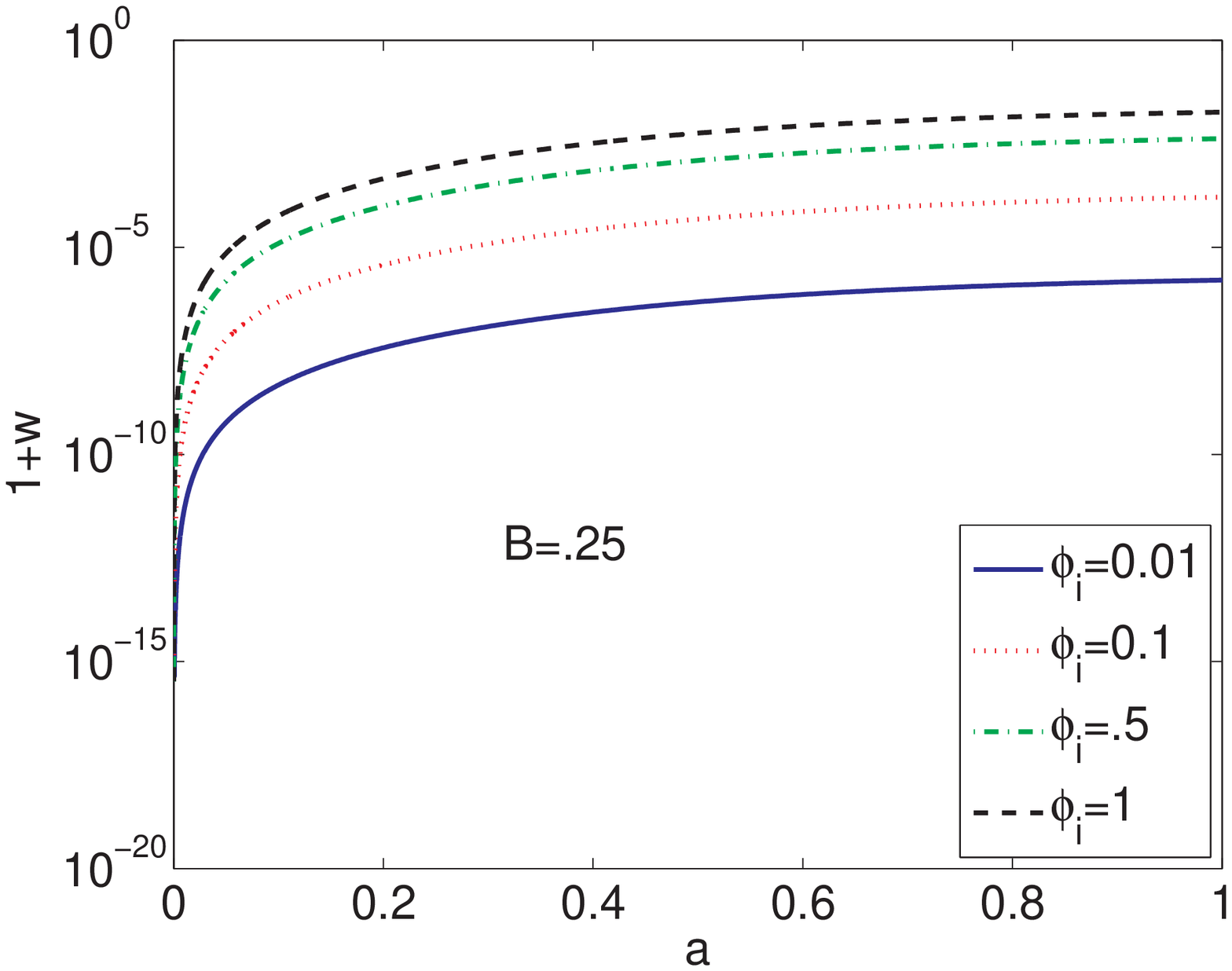,height=51mm}
	\caption
	{	\label{onepw_vs_a_B0p25} $1+w(a)$ for different choices of $\phi_i$ (on a logarithmic scale). $B$ is fixed to $0.25$}
\end{figure}

\fig{onepw_vs_a_B0p25} shows (on a logarithmic scale) the effect of varying the initial condition on $\phi$ for a given choice of $B=0.25$. Note that in our units, $\phi$ is measured in units of the Planck Mass $M_{\rm Pl}$. We find that small values of $\phi_i$ lead to a very slow evolution of $w$, whereas large values lead to a stronger fluctuation in $w$. If one makes the same plots for smaller (or larger) fixed values of $B$, the shapes of the plots remain the same, though the amplitude of fluctuation of $w$ is smaller (or larger).

As explained in the next section, the parameter choices $(B,\phi_i)=(.5,1)$, correspond to the upper limits of the ranges of $B$ and $\phi_i$ explored in this paper. From \fig{onepw_vs_a_phi1} we find that the final $1+w$ in this extreme case is only about $0.06$. This shows that our slow-roll approximation is valid in all cases examined.

\section{Comparison to Observations}
In this section, we use observational data from Type Ia supernovae, Baryon Acoustic Oscillations and the Cosmic Microwave Background to constrain the three parameters $\Omega_{m0}$, $B$ and $\phi_i$.

\paragraph{Type Ia Supernovae constraints}
To place standard candle constraints on the model, we use the recently released Union08 compilation of SnIa data \cite{union08}. This is a heterogeneous
dataset consisting of data from the Supernova Legacy Survey, the Essence survey, the recently extended dataset of distant supernovae observed with the Hubble Space Telescope, as well as older datasets.

The $\chi^2$ from SNIa is calculated as follows:
\be
\chi ^2 _{SN}  = \frac{{\sum\limits_{i = 1}^N {\left[ {\mu _{\text{obs} } \left( {z_i } \right) - \mu _{\rm th} \left( {z_i } \right)} \right]} ^2 }}{{\sigma^{2} _{\mu,i} }}
\ee
where $N=307$ is the number of SNIa data points. $\mu_{\rm obs}$ is the  observed distance modulus, defined as the difference between the apparent and absolute magnitude of the supernova. The $\sigma_{\mu,i}$ are the errors in the observed distance moduli, arising from a variety of sources, and assumed to be gaussian and uncorrelated. The theoretical distance modulus $\mu_{\rm th}$  depends on the model parameters via the dimensionless luminosity distance $D_{L}(z)$:
\be
D_{L}\(z\)\equiv\left(1+z\right) \int^{z}_{0}dz'\frac{H_0}{H\left(z';\Omega_{m0},B,\phi_{i}\right)}
\ee
as follows:
\be
\mu_{\rm th}\left(z\right)=42.38-5\log_{10}h+5\log_{10}\[D_{L}\left(z\right)\]
\ee

We marginalize over the present value of the Hubble parameter following the techniques described in \cite{perivol1} and construct $\chi^2$ likelihood contours for the parameters $\Omega_{m0}$ and $B$ for different choices of $\phi_i$.

\paragraph{CMB constraints}
We use the CMB data to place constraints on the parameter space following the recipe described in \cite{Komatsu:2008hk}. The ``CMB shift parameters'' \cite{Wang1,Wang2} are defined as follows:
\be
R\equiv \sqrt{\Omega_m\(0\)}H_0 r\(z_*\),\,\quad l_{a}\equiv \pi r\(z_*\)/r_{s}\(z_*\)
\ee
Here $r(z)$ is the comoving distance to redshift $z$ defined as:
\be
r(z)\equiv\int_{0}^{z}\frac{1}{H\(z\)}dz
\ee
$r_{s}\(z_*\)$ is the comoving sound horizon at decoupling (redshift $z_*$) given by
\be
r_{s}\(z_*\)=\int_{z_*}^{\infty}\frac{1}{H\(z\)\sqrt{3\(1+R_{b}/\(1+z\)    \)}}dz
\ee
The quantity $R_b$ is the photon-baryon energy-density ratio, and its value can be calculated as $R_b=31500 \Omega_{b} h^2 \(T_{CMB}/2.7K\)^{-4}$. The redshift at decoupling $z_*\(\Omega_b,\Omega_m,h\)$ can be calculated from the formulas in \cite{husugiyama}.

$R$ can be physically interpreted as a scaled distance to recombination, and $l_{a}$ can be interpreted as the angular scale of the sound horizon at recombination.

The $\chi^2$ contribution of the CMB is given by
\be
\chi^{2}_{CMB}=\mathbf{V}_{\rm CMB}^{\mathbf{T}}\mathbf{C}_{\rm inv}\mathbf{V}_{\rm CMB}
\ee
Here $\mathbf{V}_{\rm CMB}\equiv\mathbf{P}-\mathbf{P}_{\rm data}$, where $\mathbf{P}$ is the vector $\(l_{a},R,z_{*}\)$ and the vector $\mathbf{P}_{\rm data}$ is formed from the WMAP $5$-year maximum likelihood values of these quantities \cite{Komatsu:2008hk}. The inverse covariance matrix $\mathbf{C}_{\rm inv}$ is also provided in \cite{Komatsu:2008hk}.

\paragraph{Baryon Acoustic Oscillation constraints}
Finally, we also use the Baryon Acoustic Oscillations (BAO) to constrain the model.  The measured quantity here is the ratio $r_{s}\(z_{*}\)/D_{V}\(z\)$, where $D_{V}\(z\)$ is the so called ``volume distance'' defined in terms of the angular diameter distance $D_{A}\equiv r\(z\) /\(1+z\)$ as
\be
D_{v}\(z\)\equiv\left[\frac{\(1+z\)^2 D_{A}^{2}(z) z }{H(z)}\right]^{1/3}
\ee
So far the BAO peak has been measured at two redshifts, $z=0.2$ and $z=0.35$ \cite{Seo:2005ys,Percival:2007yw}. The ratio of the two measurements of $D_{v}\(z\)$, i.e., $D_{v}\(.35\)/D_{v}\(.2\)=1.812\pm0.060$ \cite{Percival:2007yw}, can be used as a model-independent observational constraint. In this paper, following \cite{perivol1}, we calculate the $\chi^2$ contribution of the BAO measurements as follows:

\be
\chi^{2}_{BAO}=\mathbf{V}_{\rm BAO}^{\mathbf{T}}\mathbf{C}_{\rm inv}\mathbf{V}_{\rm BAO}
\ee

The vector $\mathbf{V}_{\rm BAO}\equiv\mathbf{P}-\mathbf{P}_{\rm data}$, with $\mathbf{P}\equiv \( D_{v}\(0.32\),D_{v}\(0.2\) \) $ and $\mathbf{P}_{\rm data}\equiv\(0.1980, 0.1094\)$, the two measured BAO data points  \cite{Percival:2007yw}. The inverse covariance matrix is provided in \cite{Percival:2007yw}.

\begin{figure}
	\epsfig{file=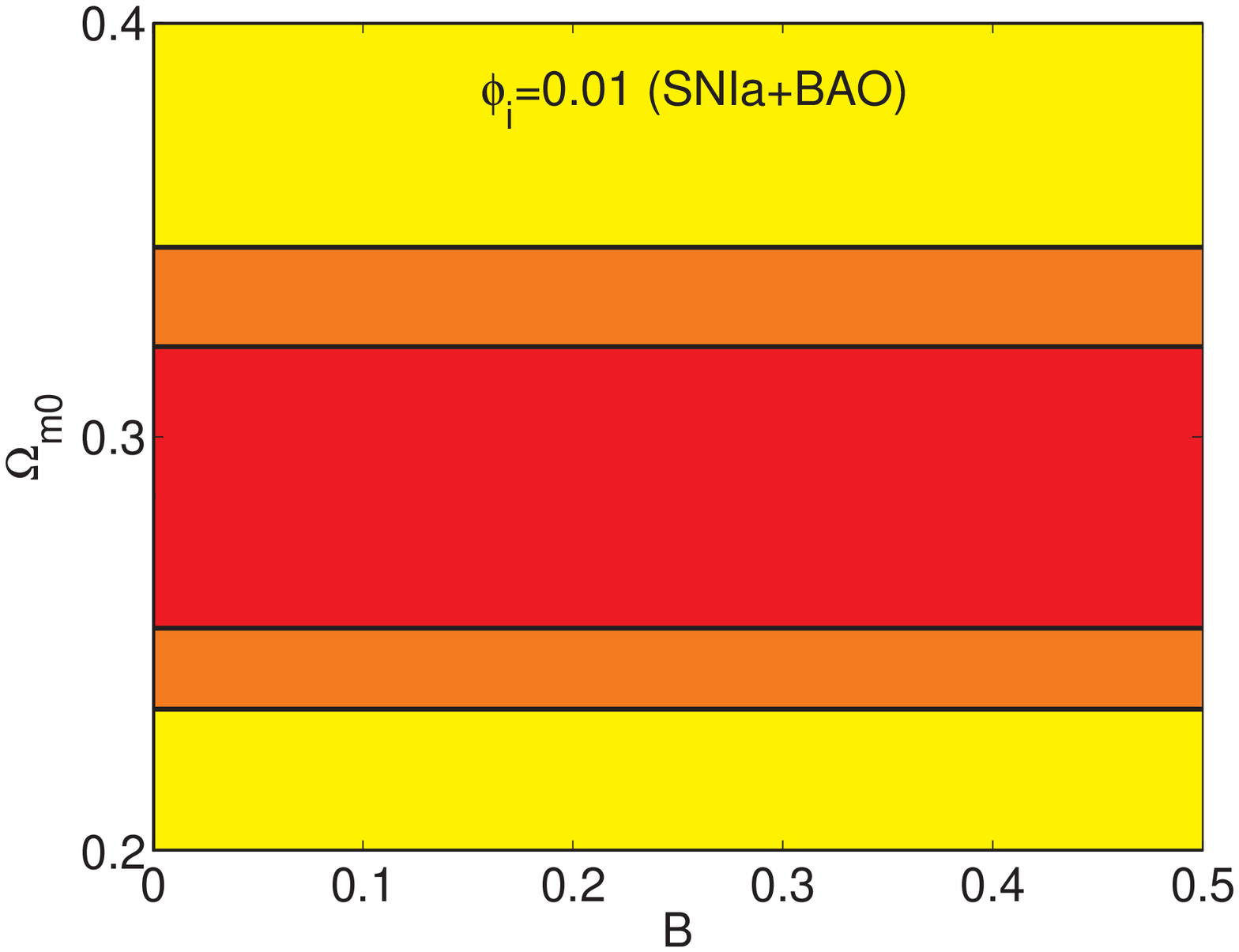,height=51mm}
	\caption
	{	\label{SNBAO_phi0p01} Likelihood plot from SNIa+BAO data for the parameters $B$ and $\Omega_{\Lambda}$ with $\phi_i=0.01$.
The yellow (light) region
is excluded at the 2$\sigma$ level, and the darker (orange) region
is excluded at the 1$\sigma$ level.  Red (darkest) region is
not excluded at either confidence level.}
\end{figure}

\begin{figure}
	\epsfig{file=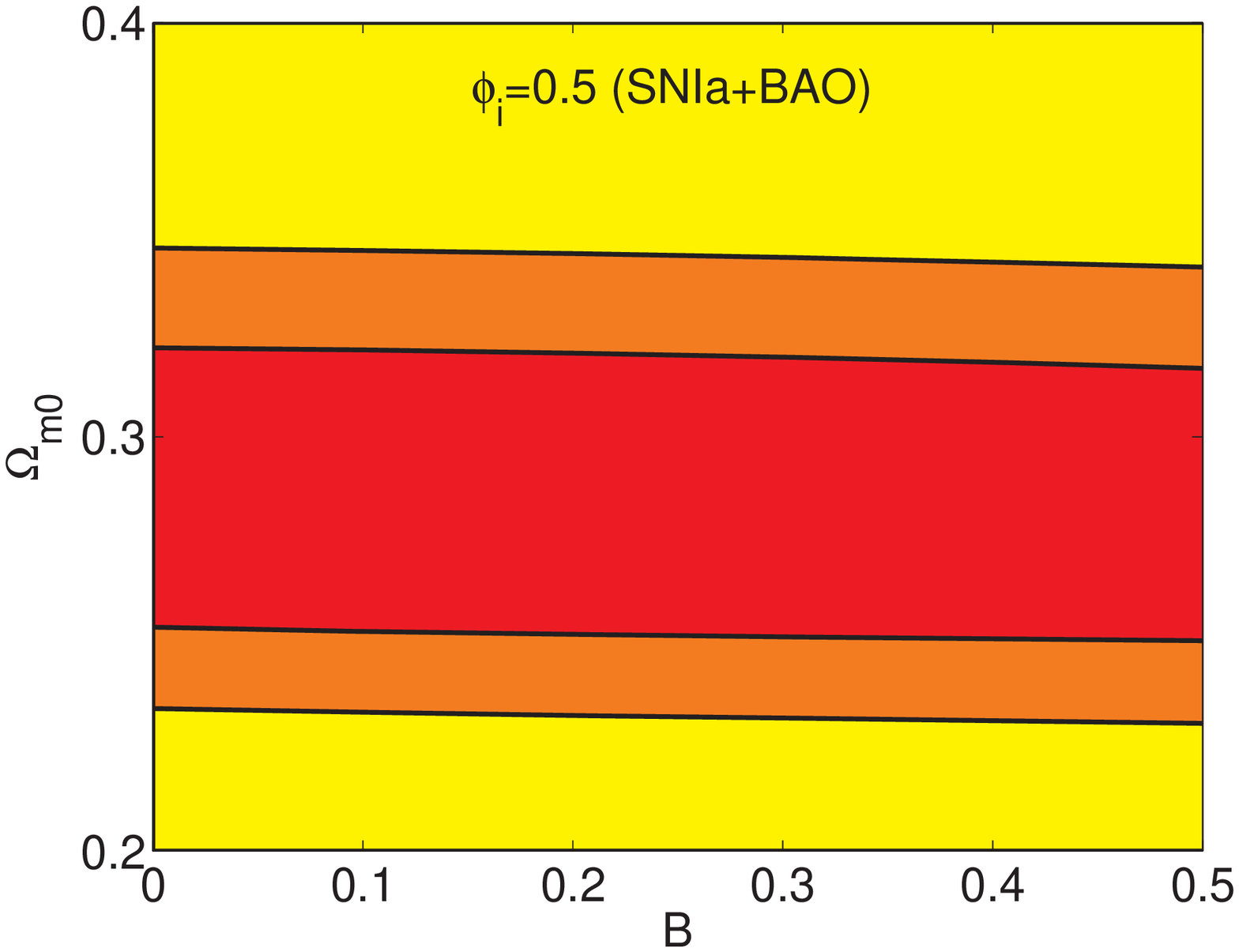,height=51mm}
	\caption
	{	\label{SNBAO_phi0p5}As for \fig{SNBAO_phi0p01} with $\phi_i=0.5$.
}
\end{figure}

\begin{figure}
	\epsfig{file=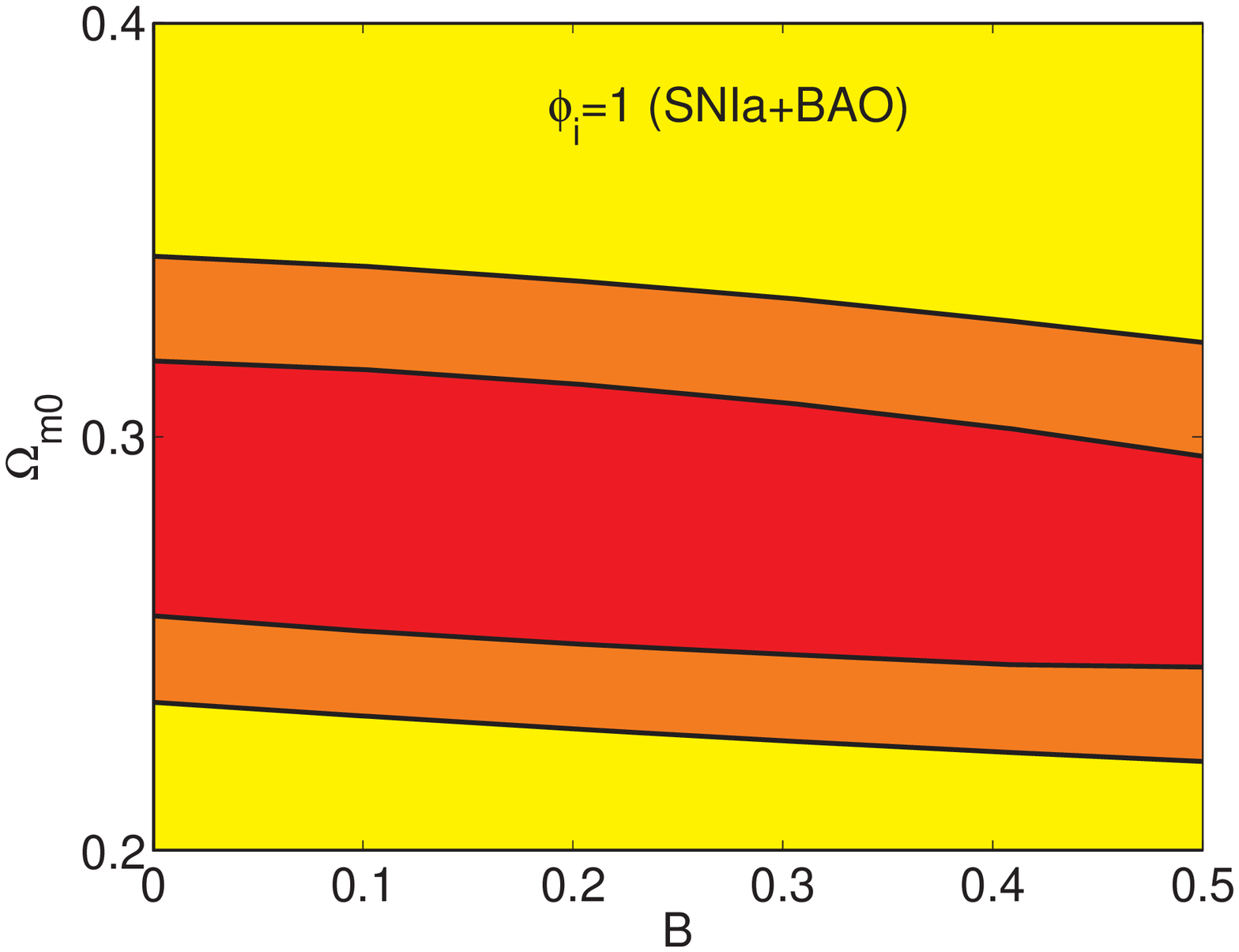,height=51mm}
	\caption
	{	\label{SNBAO_phi1}As for \fig{SNBAO_phi0p01} with $\phi_i=1$}
\end{figure}

\paragraph{Constraint contours}
In figures \ref{SNBAO_phi0p01}-\ref{SNBAO_phi1} we construct joint SNIa+BAO constraint contours for the variables $\Omega_{\phi 0}$ and $B$ for three different choices of $\phi_i$, marginalizing over the present day Hubble parameter $h$.

For the parameter $B$, we examine the range $0\leq B\leq 0.5$. This is in accordance with \eqn{Bdu}, using the fact that $1<d_u<2$ \cite{Georgi:2007si}. However, it is important to remember that given our assumptions, \eqn{Bdu} is true only up to an overall O(1) factor.  The limiting case of $B=0$ corresponds to a $\Lambda$CDM cosmology.

\begin{figure}
	\epsfig{file=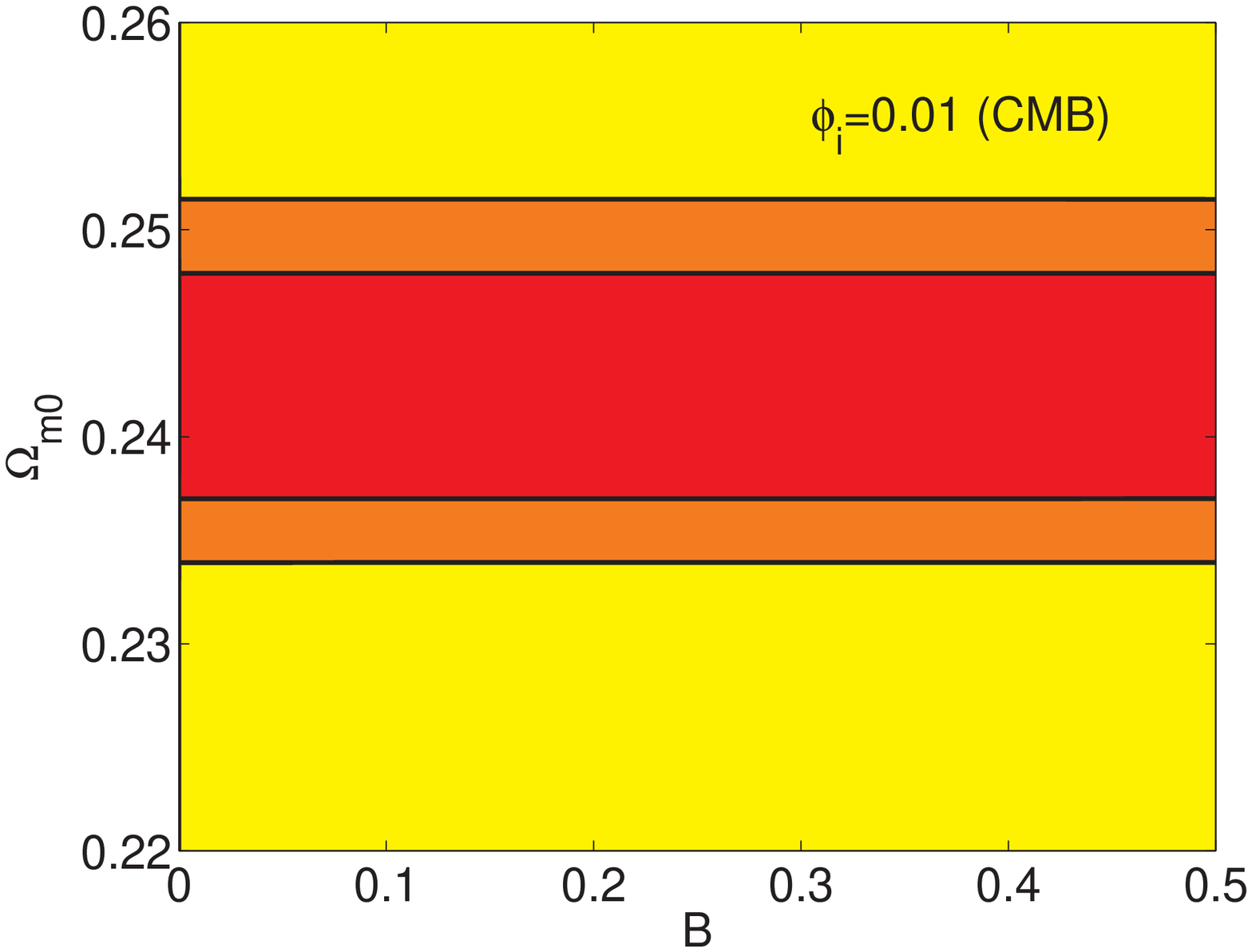,height=51mm}
	\caption
	{	\label{CMB_phi0p01}Likelihood plot from CMB data for the parameters $B$ and $\Omega_{\Lambda}$ with $\phi_i=0.01$.
The yellow (light) region
is excluded at the 2$\sigma$ level, and the darker (orange) region
is excluded at the 1$\sigma$ level.  Red (darkest) region is
not excluded at either confidence level.}
\end{figure}

\begin{figure}
	\epsfig{file=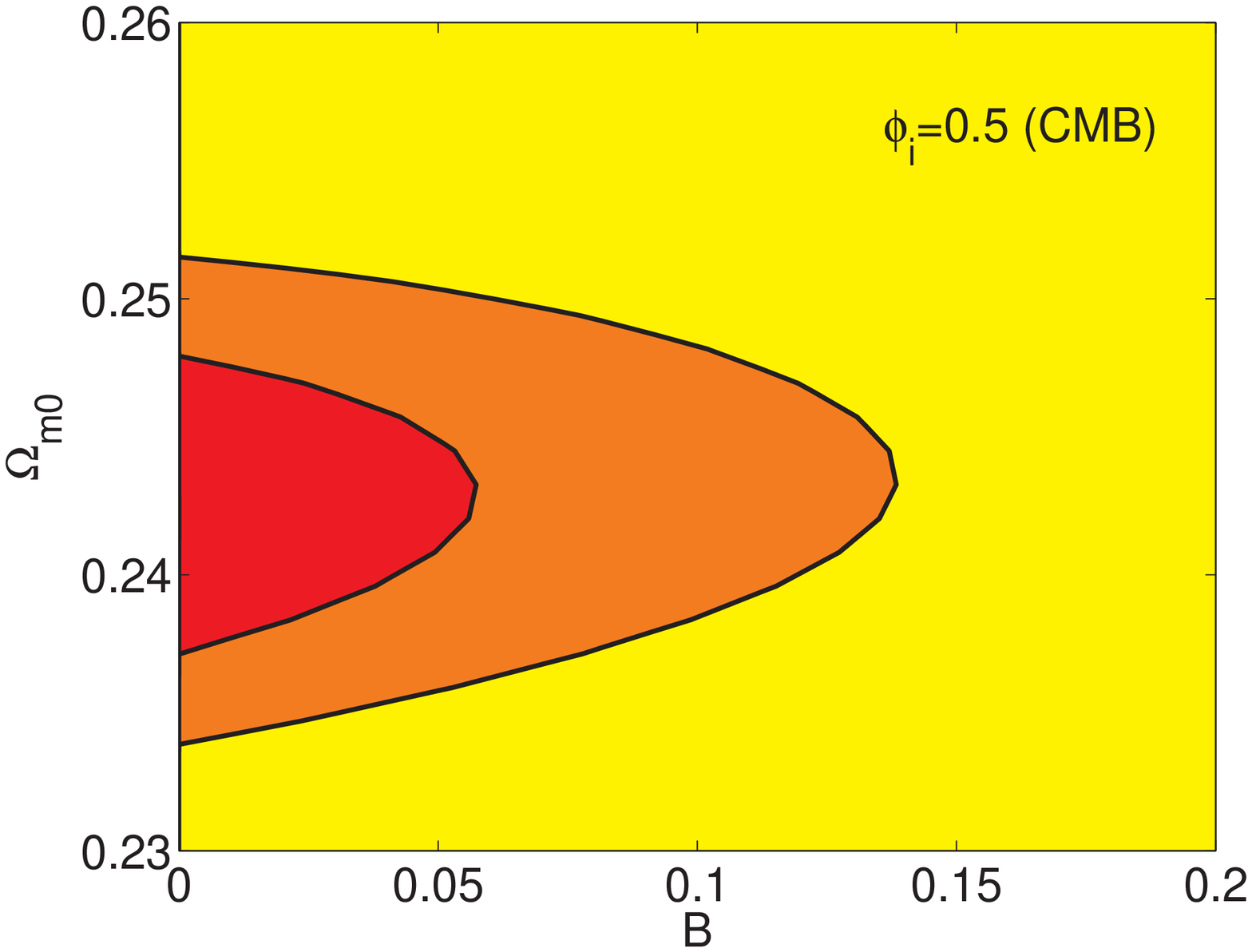,height=51mm}
	\caption
	{	\label{CMB_phi0p5}As for \fig{CMB_phi0p01} with $\phi_i=0.5$}
\end{figure}

\begin{figure}
	\epsfig{file=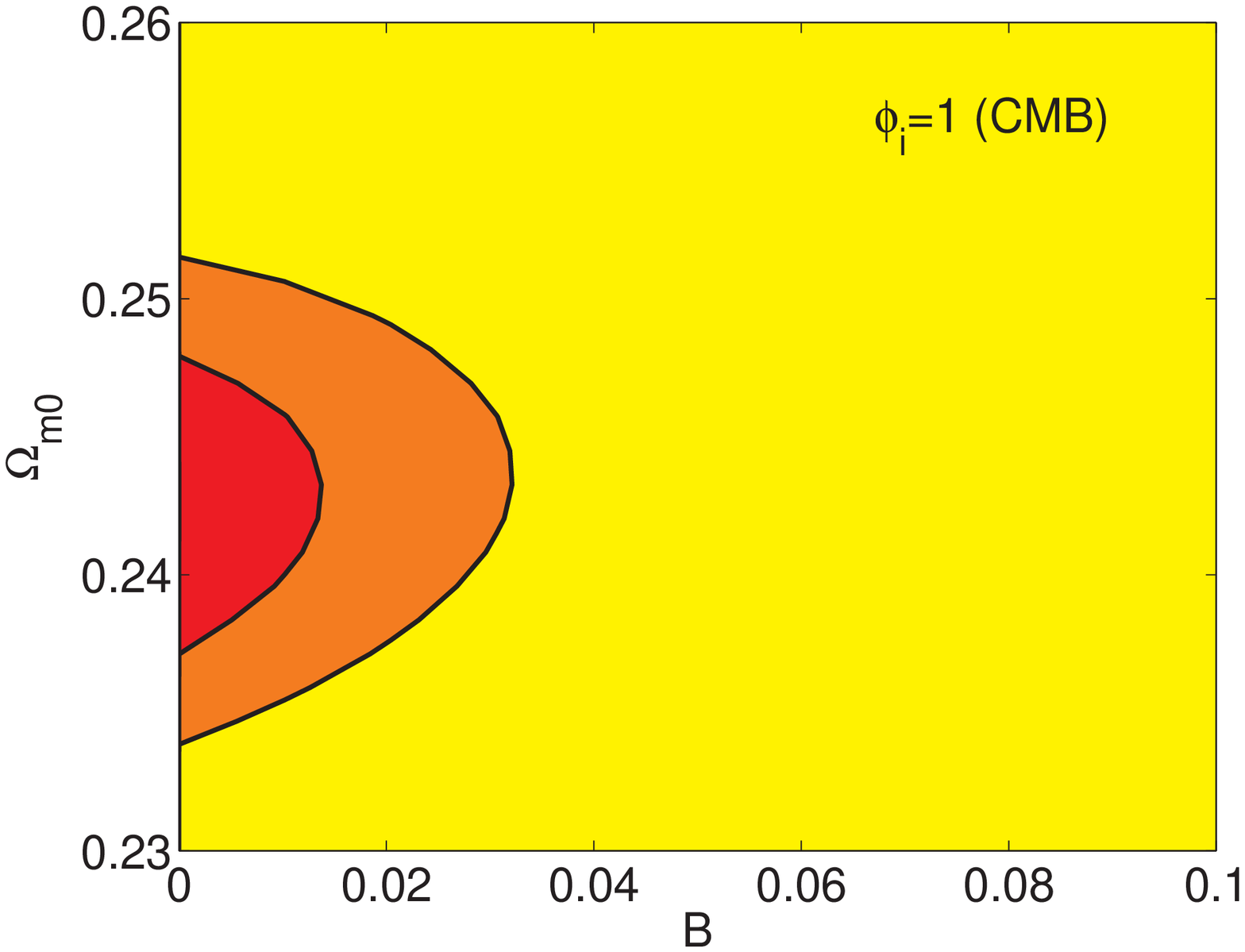,height=51mm}
	\caption
	{	\label{CMB_phi1}As for \fig{CMB_phi0p01} with $\phi_i=1$}
\end{figure}

In figures \ref{CMB_phi0p01}-\ref{CMB_phi1} we construct $(\Omega_{\phi 0},B)$ contours for three different values of $\phi_i$ (or equivalently, $\lambda$), $\phi_i={.01,0.5,1}$ using the CMB data. Recall that in our units, $\phi$ is measured in units of $M_{\rm Pl}$, and hence the extreme case of $\phi_i=1$ corresponds to the maximum unparticle mass being on order of the Planck mass.

 Clearly, the observational data do not rule out these models. We find that the contours slightly shrink as $\phi_i$ increases. However, $B$ is poorly constrained by the SnIa+BAO data for all choices of $\phi_i$ examined, as well as the CMB data for small $\phi_i$. This is because a smaller choice of $\phi_i$ leads to a slower evolution of the field for all choices of $B$ (within the range examined). For large values of $\phi_i$ (closer to the Planck Mass), the CMB data place stringent constraints on $B$.

% \textcolor{green}{????????}
%\textcolor{blue}{ The constraint contours imply that the current cosmological data are consistent with the models considered as long as $\lambda\sim 10^{-2}M_{\rm Pl}$ or smaller, for essentially the entire range of $1<d_u<2$. For larger $\lambda$, the range of $d_u$ that the data is consistent with is smaller. Phenomenologically interesting cases, $\lambda\sim$TeV are therefore consistent with cosmological data.}

\section{Conclusions}
We have studied model of dark energy in which an unparticle degree of freedom plays the role of a slowly rolling scalar field quintessence. While fundamental scalar fields are not protected from radiative corrections, unparticles are, since they are bound states of  fundamental fermions whose mass is protected. [We note however that a proper analysis of quantum corrections in the unparticle context would be highly useful.] Further, coupling of unparticles to the standard model fields can be arbitrarily suppressed by raising the fundamental energy scale $M_F$. Therefore, the unparticle dark energy model is free of most of the problems that plague fundamental scalar field quintessence models.

We examined how the dynamics of the unparticle field equation of state is determined by the parameters $\Omega_{m,0}$, the constant $B$ (a function of the unparticle scaling dimension $d_u$), and the initial value of the field $\phi_i$ (which we assume to be comparable to the unparticle energy scale $\lambda_u$).  We have used the SNIa, CMB and BAO to place observational constraints on these parameters and find that such models are not ruled out by the data.

\section{Acknowledgments}
S.D. was supported in part by the Department of Energy (DE-FG05-85ER40226). D.S. acknowledges the financial support from NSF, award number PHY-0914893.

\end{document}